\documentclass[aps,showkeys,twocolumn,superscriptaddress]{revtex4}
\usepackage{graphicx}
\usepackage{amsmath}
\usepackage{subfigure}
\usepackage{lipsum}
\usepackage{color}

\begin{document}

\renewcommand{\c}{\textcolor{red}{[ref]}}

\title{Graphene infrared light emitting diode (GILED)}

\author{Z.Z. Alisultanov\footnote{Corresponding author.\\Electronic address: zaur0102@gmail.com}}
\affiliation{Amirkhanov Institute of Physics Russian Academy of Sciences, Dagestan Science Centre, Makhachkala, Russia}
\affiliation{Prokhorov General Physics Institute Russian Academy of Sciences, Moscow, Russia}
\affiliation{Dagestan State University, Makhachkala, Russia}

\author{M.S. Reis}
\affiliation{Instituto de F\'isica, Universidade Federal Fluminense, 24210-346, Niter\'oi-RJ, Brazil}

\keywords{graphene, optoelectronics, bandgap, LED}

\date{\today}

\begin{abstract}
The present Letter proposes a device based on graphene for infrared light emission. It is based on a n- and p-doped monolayer graphene (MGs), with Fermi energies $E_F$ and -$E_F$, respectively, sandwiching a bilayer graphene (BG) with bandgap $\Lambda=2|eV_g-\Delta|\geq 2E_F$, where $V_g$ is the gate voltage across the BG and $\Delta$ the sub-lattice energy difference into each layer of the BG. This device works as simple as tuning the gate voltage to decrease the BG bandgap down to $2E_F$; and, once this condition is fulfilled, a current flows from the n-doped MG to the p-doped MG. However, when electrons achieve the other side of the device, i.e., into the p-doped MG, their energies ($E_F$) are much bigger than the holes energies ($-E_F$), and thus these electrons decay emitting infrared photons.
\end{abstract}

\maketitle

Graphene has an unique electronic structure, with a conical conduction and valence bands converging to the Dirac point; and, as a consequence, electrons of this material behave as massless Dirac fermions. This feature gives rise to a number of phenomena, such as: quantum conductances\cite{novoselov2005two,balandin2008superior}; frequency-independent optical conductance for a wide range of photons energy and, consequently, unique optical transmittance and broadband absorption\cite{bao2012graphene}; quantum Hall effect\cite{zhang2005experimental}; high carrier mobility\cite{bolotin2008ultrahigh}; optical transparency\cite{nair2008fine}; mechanical flexibility\cite{dikin2007preparation}; environmental stability\cite{bonaccorso2010graphene} and peculiar thermodynamic oscillations\cite{alisultanov2016quantum,alisultanov2016magneto,reis2013electrocaloric,reis2012oscillating}. 

Graphene is therefore a truly example of advanced functional material, and the most potential interface for graphene applications lies between optoelectronics and photonics, where those unique electronic and optical properties can be maximally and completely explored. Within this scenario we are able to cite applications on ultrafast lasers\cite{otsuji2013terahertz}, touch screens\cite{bonaccorso2010graphene}, photodetectors\cite{xia2009ultrafast}, light-emitting structures\cite{matyba2010graphene} and many other devices; as well as graphene based devices into the THz region, that lies between microwaves zone and infrared-visible region, both zones with efficient and well established devices already\cite{zhou2014graphene}. In what concerns light-emitting devices based on graphene, it has been proposed, for instance, that few layers graphene can replace ITO (indium tin oxide), or any other transparent conductors, as OLED anodes\cite{bao2012graphene,matyba2010graphene}. The present Letter goes therefore on this direction and takes advantage of the opto-electronic properties of graphene to propose the infrared light emitting diode (GILED).

However, for a reliable graphene application on opto-electronics and photonics the central role for an efficient device is the ability to manage the graphene bandgap.  For a monolayer graphene (MG), it is possible to induce a tunable bandgap by applying either a vertical electric field or a quantum confinement, like in graphene nanoribbons (GNR).  The former case displays a bandgap of up to 0.25 eV for an electric field of 3 V/nm \cite{zhang2009direct}; while the second case presents a bandgap of up to 1 eV for a GNR of few nm width. This last case has an extra advantage to be easily integrated in nanoelectronics due to its reduced dimensions. A detailed review of bandgap engineering has been published by Lam and Guo\cite{lam2014bandgap} and, in addition to the electric field and quantum confinement, other methods are still on the focus of the scientific community, as mechanical stress and chemical modification of the graphene composition. This bandgap engineering ensures to graphene a wide sort of nano-opticalelectronic applications. 
\begin{figure*}[t!]
\begin{center}
\subfigure{\includegraphics[width=5.5cm]{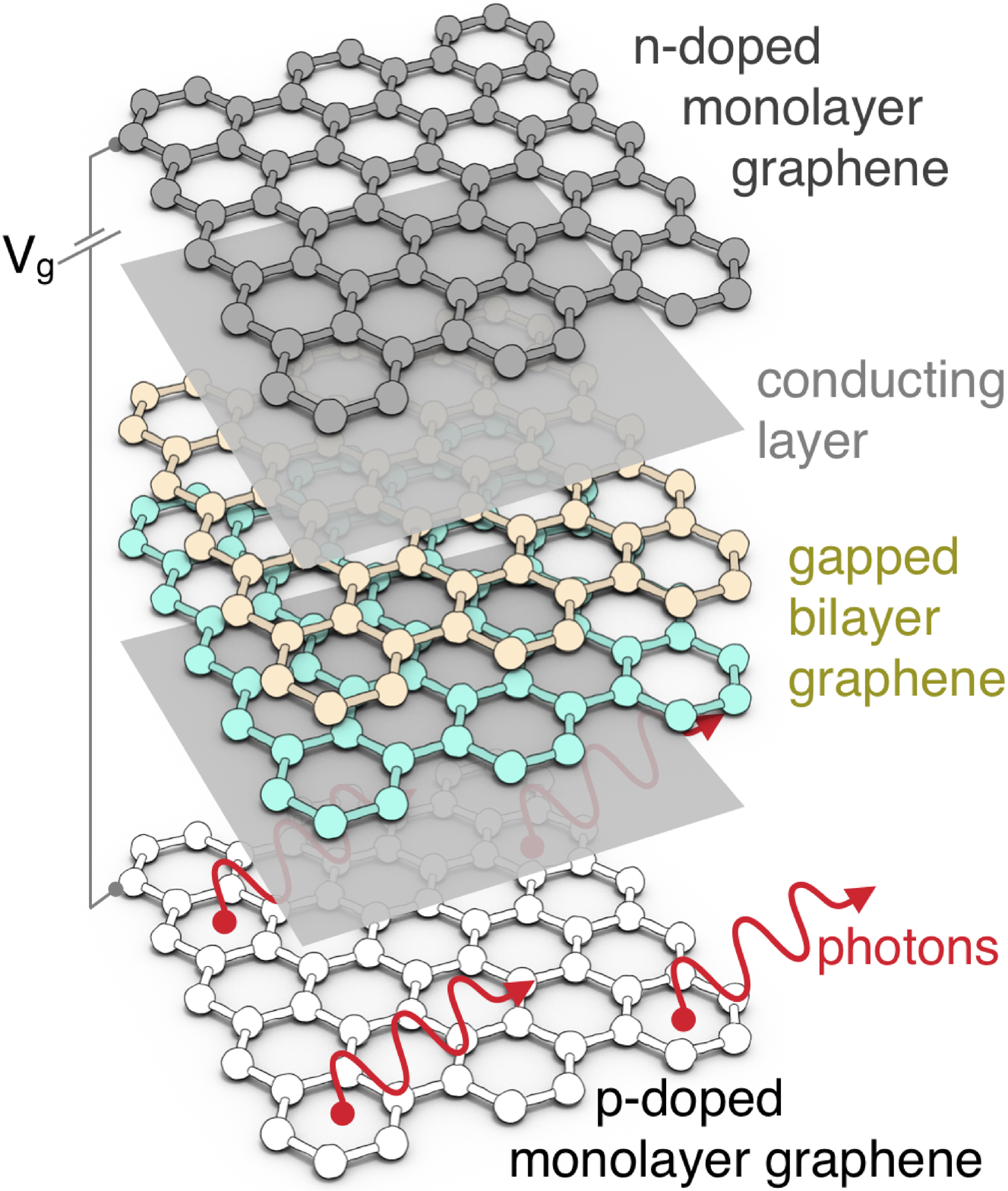}}\hfill
\subfigure{\includegraphics[width=5.5cm]{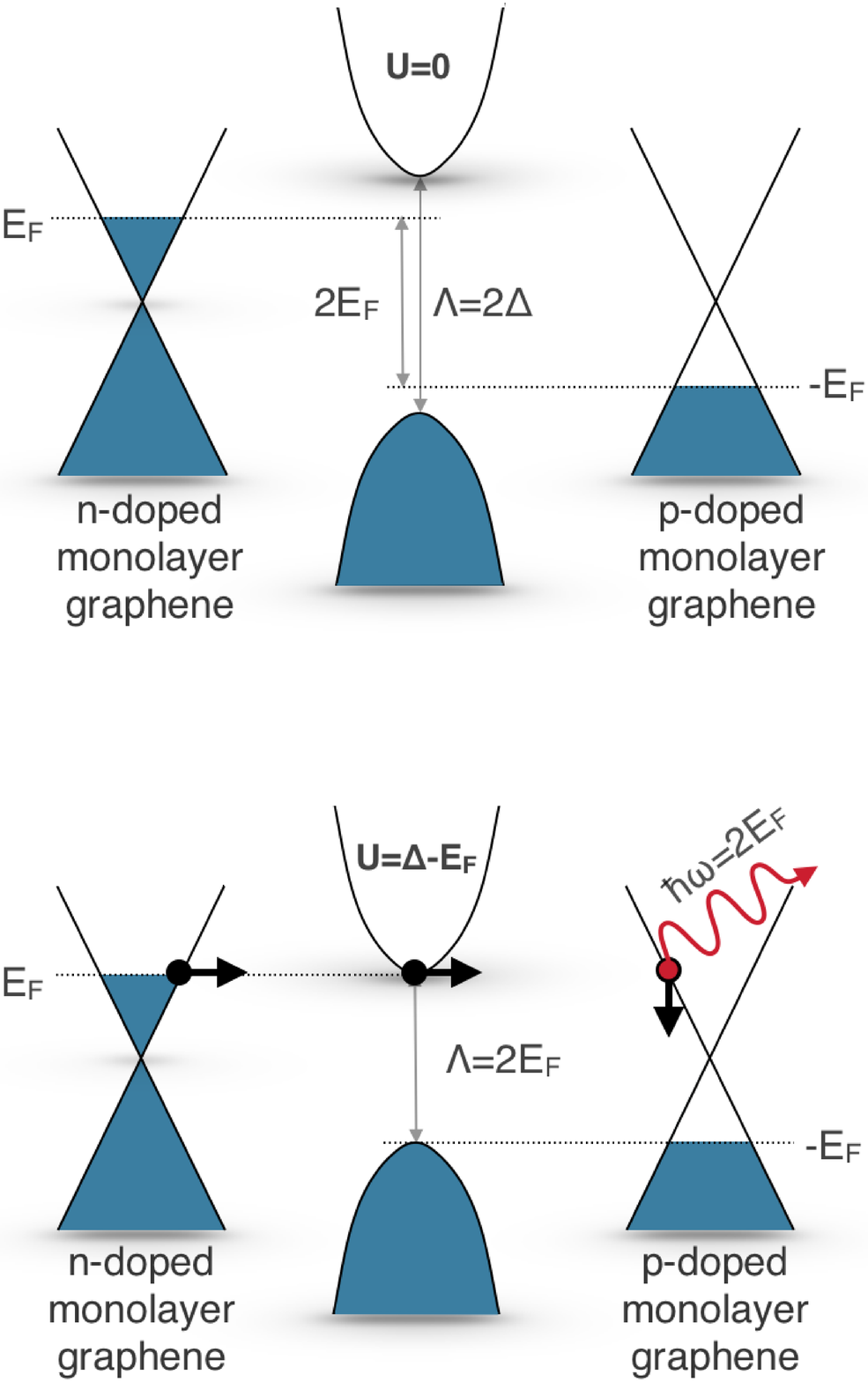}}\hfill
\subfigure{\includegraphics[width=5.5cm]{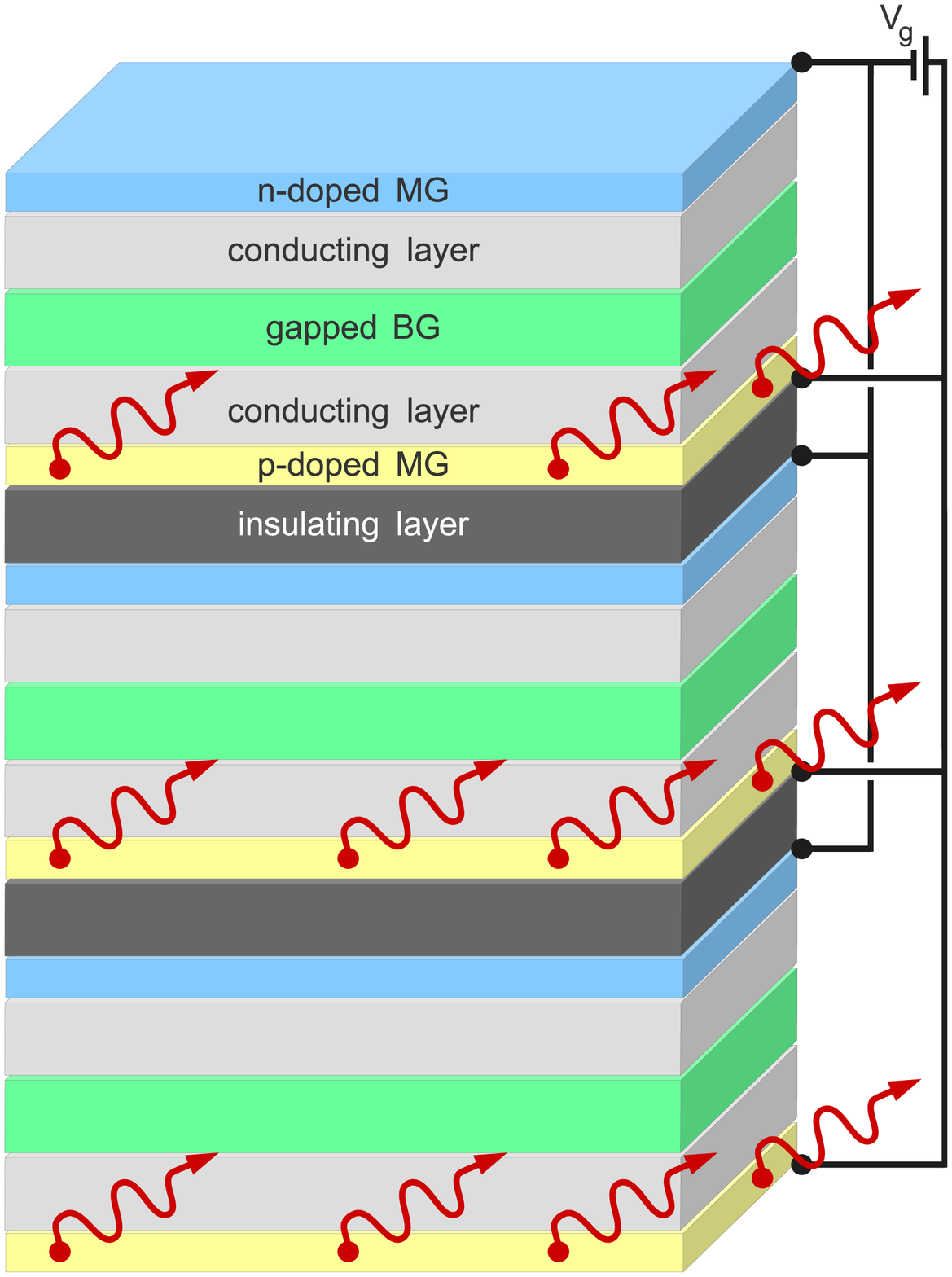}}
\end{center}
\caption{(color online) Left: GILED conception: a n- and p-dopped MGs sandwiching a gapped BG. Middle-top: band structure of the MGs and the gaped BG for zero gate voltage, i.e., device turned off. There is no current flow trough the device, since $\Delta>E_F$. Middle-bottom: band structure of the MGs and the gaped BG for a gate voltage that fulfills the condition $eV_g=\Delta-E_F$ (and, consequently, $\Lambda=2E_F$), i.e., device turned on. This case allows a current flow trough the device; and then, when electrons achieve the p-doped MG, these decay emitting infrared photons. Right: Parallel association of some GILEDs as a multilayer device. This structure could increase the emission of photons.\label{device}}
\end{figure*}

The above description is not limited to MG, and bilayer graphene (BG) has also received the due attention\cite{lam2014bandgap}. It is constituted by two MGs stacked in some possible ways as, for instance, the Bernal packing\cite{katsnelson2012graphene}, that leads to a quadratic dispersion relation $E=\pm\hbar^2k^2/2m$, but without bandgap. However, considering the sub-lattice energy difference $\Delta$ into each layer of the BG, the dispersion relation assumes a more sophisticated structure\cite{alisultanov2016electron,mucha2010electron,mccann2013electronic}:
\begin{equation}
E=\pm\left\{\Delta^2+|f(\vec{k})|^2+\frac{t_\perp^2}{2}\pm t_\perp\left[|f(\vec{k})|^2+\frac{t_\perp^2}{4}\right]^{1/2}\right\}^{1/2}
\end{equation}
where $t_\perp$ is the hopping energy between the two layers, 
\begin{equation}
|f(\vec{k})|^2=3+2\cos\left(\sqrt{3}k_xa\right)+4\cos\left(\sqrt{3}k_xa/2\right)\cos\left(3k_ya/2\right),
\end{equation}
$a\approx1.42$ \AA is the carbon-carbon distance, $\vec{k}=\vec{q}-\vec{Q}$, $\vec{q}$ is the two dimensional wave vector of graphene's electrons and $\vec{Q}$ stands for the Dirac point coordinate in momentum space. For this case, a bandgap appears with width of $\Lambda=2\Delta$\cite{alisultanov2016electron,mucha2010electron,mccann2013electronic}, but it is not possible to manage its value by means of any external parameter. However, when the BG is under a gate voltage $V_g$ (with a corresponding potential $U=eV_g$), the energy spectrum resumes as \cite{alisultanov2016electron}:
\begin{align}\label{disper_com}
E=&\pm\left\{U^2+\Delta^2+|f(\vec{k})|^2+\frac{t_\perp^2}{2}\pm\right. \\\nonumber
&\left.\left[(t_\perp^2+4U^2)|f(\vec{k})|^2+4U^2\Delta^2+2t_\perp^2\Delta U+\frac{t_\perp^4}{4}\right]^{1/2}\right\}^{1/2}
\end{align}
and, consequently, the bandgap turns to be:
\begin{equation}\label{lambda_final}
\Lambda=2\left[\Delta^2+U^2-\frac{2}{t_\perp^2+4U^2}\left(2U^4+2U^2\Delta^2+t_\perp^2\Delta U\right)\right]^{1/2}
\end{equation}
Naturally, $k=0$ leads to $\Lambda=2|U-\Delta|$ and \textit{this is the key result to enable the GILED device}, since now it is possible to manage the bandgap width by means of an external parameter; and, for this case, it is the gate voltage. Note, the bandgap can either increase or decrease by changing the gate voltage. The dispersion relation near the Dirac point (equation \ref{disper_com}), has been described in details on reference \cite{alisultanov2016electron}. In fact, the bandgap for $U=0$ is 2$\Delta$; and increasing the gate voltage the bandgap decreases down to zero, at $U=\Delta$. Further increasing of $U$ (higher than $\Delta$), differently as before, promotes an increasing of the bandgap. More precisely (see reference \onlinecite{alisultanov2016electron}), for $U<\Delta$, the gap exists and the spectrum has a parabolic profile; while for $U>\Delta$, the gap still existes and the energy spectrum assumes a `Mexican hat' shape. 

These results consider the bandgap caused by a violation of the equivalence between sublattices into layers of the BG; however, a similar result can be written for the case when the bandgap is caused by the energy difference $\Delta^\prime$ between layers of the BG. Technically speaking, it is easier to be achieved in comparison the the former case, since a simple adsorption of atoms on one of the layers of the BG\cite{ohta2006controlling} can add some extra Coulomb energy to the system, favoring the violation of symmetry. For this case, the bandgap dependence on the gate voltage has a behavior qualitatively the same as the one described above.

The GILED conception considers a n-dopped MG electronically linked to a gapped BG under a gate voltage that, on its turn, is also electronically linked to a p-dopped MG, as shown on figure \ref{device}-left. The MGs are doped in such a way that the Fermi energy of the n-doped and p-doped MGs are equal $E_F$ and $-E_F$, respectively. When the device is turned off, the gate voltage is zero (and, consequently, the potential $U$ is also zero), and the bandgap $\Lambda$ satisfies the condition $\Lambda=2\Delta>2E_F$. This situation does not allow a current flow between MGs  (see figure \ref{device}-middle-top). To turn on the device, the gate voltage is turned on and thus, as can be seen on figure \ref{device}-middle-bottom, the potential $U$ ($<\Delta$) increases; and, consequently, the bandgap decreases (here is the key effect of the device, taking advantage of the possibility to decrease the bandgap by managing the gate voltage). The condition $\Lambda=2E_F$ is achieved for $U=\Delta-E_F$ and then a current flow is stablished from the n-doped MG to the p-doped MG through the BG. However, as soon as the electrons reach the p-doped MG, these decay to the valence band by emitting a photon with energy $\hbar\omega=2E_F$ (see figure \ref{device}-middle-bottom). This system behaves similarly to a LED; but the mechanism of action of this GILED is fundamentally different from conventional LEDs. 

n- and p-doped graphene can be obtained and tuned by, for instance, ion and molecule doping of the MG sheet\cite{guo2010controllable,dai2010modulating}; and typical value of the Fermi energy is $90$ meV\cite{raza2012graphene}. For this case,  the emitted photon has c.a. 45 THz and therefore within the infra-red region. The potential gate $V_g$ to achieve this emission is simply c.a. 10 mV, considering $\Delta\approx100$ meV\cite{mccann2013electronic,mucha2010electron}.

This device can indeed be enhanced by developing a parallel chain of GILEDs, increasing the flux of emitted photons. In this case, a multilayer of single devices could be built, separated by an insulating sheet. This proposed multilayer device is depicted on figure \ref{device}-right.

Summarizing, this Letter proposes a graphene based infrared light emitting diode (GILED). This device takes advantage of the possibility to tune the bandgap of a BG due to a voltage gate across the layers and works as simple as a n-doped MG/gaped BG/p-doped MG sandwich. More precisely, the Fermi energy of the MGs is bigger than the bandgap of the BG and for a certain value of the gate voltage ($eV_g=\Delta-E_F$), a current channel thought the device is opened; i.e., electrons from the conduct band of the n-doped MG flow trough the BG and achieve the p-doped MG. However, the energy of these electrons ($E_F$) are much bigger than the Fermi energy of the p-doped MG ($-E_F$), and then these decay by emitting photons with energy $2E_F$. This proposal is purely theoretical and a proof-of-concept realization of this device is highly welcomed.

\begin{acknowledgments}
ZZA declares that this work was supported by grants Nos. RFBR 15-02-03311a, PG MK-4471.2015.2, RSCF MKH-15-19-10049 and 3.1262.2014 from the Ministry of Education and Science of Russia.  ZZA is also sincerely grateful to the Dmitry Zimin Foundation ÒDynastyÓ for
financial support. MSR acknowledges FAPERJ, CAPES, CNPq and PROPPI-UFF for financial support. 
\end{acknowledgments}


\begin{thebibliography}{26}
\expandafter\ifx\csname natexlab\endcsname\relax\def\natexlab#1{#1}\fi
\expandafter\ifx\csname bibnamefont\endcsname\relax
  \def\bibnamefont#1{#1}\fi
\expandafter\ifx\csname bibfnamefont\endcsname\relax
  \def\bibfnamefont#1{#1}\fi
\expandafter\ifx\csname citenamefont\endcsname\relax
  \def\citenamefont#1{#1}\fi
\expandafter\ifx\csname url\endcsname\relax
  \def\url#1{\texttt{#1}}\fi
\expandafter\ifx\csname urlprefix\endcsname\relax\def\urlprefix{URL }\fi
\providecommand{\bibinfo}[2]{#2}
\providecommand{\eprint}[2][]{\url{#2}}

\bibitem[{\citenamefont{Novoselov et~al.}(2005)\citenamefont{Novoselov, Geim,
  Morozov, Jiang, Katsnelson, Grigorieva, Dubonos, and
  Firsov}}]{novoselov2005two}
\bibinfo{author}{\bibfnamefont{K.}~\bibnamefont{Novoselov}},
  \bibinfo{author}{\bibfnamefont{A.~K.} \bibnamefont{Geim}},
  \bibinfo{author}{\bibfnamefont{S.}~\bibnamefont{Morozov}},
  \bibinfo{author}{\bibfnamefont{D.}~\bibnamefont{Jiang}},
  \bibinfo{author}{\bibfnamefont{M.}~\bibnamefont{Katsnelson}},
  \bibinfo{author}{\bibfnamefont{I.}~\bibnamefont{Grigorieva}},
  \bibinfo{author}{\bibfnamefont{S.}~\bibnamefont{Dubonos}}, \bibnamefont{and}
  \bibinfo{author}{\bibfnamefont{A.}~\bibnamefont{Firsov}},
  \bibinfo{journal}{nature} \textbf{\bibinfo{volume}{438}},
  \bibinfo{pages}{197} (\bibinfo{year}{2005}).

\bibitem[{\citenamefont{Balandin et~al.}(2008)\citenamefont{Balandin, Ghosh,
  Bao, Calizo, Teweldebrhan, Miao, and Lau}}]{balandin2008superior}
\bibinfo{author}{\bibfnamefont{A.~A.} \bibnamefont{Balandin}},
  \bibinfo{author}{\bibfnamefont{S.}~\bibnamefont{Ghosh}},
  \bibinfo{author}{\bibfnamefont{W.}~\bibnamefont{Bao}},
  \bibinfo{author}{\bibfnamefont{I.}~\bibnamefont{Calizo}},
  \bibinfo{author}{\bibfnamefont{D.}~\bibnamefont{Teweldebrhan}},
  \bibinfo{author}{\bibfnamefont{F.}~\bibnamefont{Miao}}, \bibnamefont{and}
  \bibinfo{author}{\bibfnamefont{C.~N.} \bibnamefont{Lau}},
  \bibinfo{journal}{Nano letters} \textbf{\bibinfo{volume}{8}},
  \bibinfo{pages}{902} (\bibinfo{year}{2008}).

\bibitem[{\citenamefont{Bao and Loh}(2012)}]{bao2012graphene}
\bibinfo{author}{\bibfnamefont{Q.}~\bibnamefont{Bao}} \bibnamefont{and}
  \bibinfo{author}{\bibfnamefont{K.~P.} \bibnamefont{Loh}},
  \bibinfo{journal}{ACS nano} \textbf{\bibinfo{volume}{6}},
  \bibinfo{pages}{3677} (\bibinfo{year}{2012}).

\bibitem[{\citenamefont{Zhang et~al.}(2005)\citenamefont{Zhang, Tan, Stormer,
  and Kim}}]{zhang2005experimental}
\bibinfo{author}{\bibfnamefont{Y.}~\bibnamefont{Zhang}},
  \bibinfo{author}{\bibfnamefont{Y.-W.} \bibnamefont{Tan}},
  \bibinfo{author}{\bibfnamefont{H.~L.} \bibnamefont{Stormer}},
  \bibnamefont{and} \bibinfo{author}{\bibfnamefont{P.}~\bibnamefont{Kim}},
  \bibinfo{journal}{Nature} \textbf{\bibinfo{volume}{438}},
  \bibinfo{pages}{201} (\bibinfo{year}{2005}).

\bibitem[{\citenamefont{Bolotin et~al.}(2008)\citenamefont{Bolotin, Sikes,
  Jiang, Klima, Fudenberg, Hone, Kim, and Stormer}}]{bolotin2008ultrahigh}
\bibinfo{author}{\bibfnamefont{K.~I.} \bibnamefont{Bolotin}},
  \bibinfo{author}{\bibfnamefont{K.}~\bibnamefont{Sikes}},
  \bibinfo{author}{\bibfnamefont{Z.}~\bibnamefont{Jiang}},
  \bibinfo{author}{\bibfnamefont{M.}~\bibnamefont{Klima}},
  \bibinfo{author}{\bibfnamefont{G.}~\bibnamefont{Fudenberg}},
  \bibinfo{author}{\bibfnamefont{J.}~\bibnamefont{Hone}},
  \bibinfo{author}{\bibfnamefont{P.}~\bibnamefont{Kim}}, \bibnamefont{and}
  \bibinfo{author}{\bibfnamefont{H.}~\bibnamefont{Stormer}},
  \bibinfo{journal}{Solid State Communications} \textbf{\bibinfo{volume}{146}},
  \bibinfo{pages}{351} (\bibinfo{year}{2008}).

\bibitem[{\citenamefont{Nair et~al.}(2008)\citenamefont{Nair, Blake,
  Grigorenko, Novoselov, Booth, Stauber, Peres, and Geim}}]{nair2008fine}
\bibinfo{author}{\bibfnamefont{R.~R.} \bibnamefont{Nair}},
  \bibinfo{author}{\bibfnamefont{P.}~\bibnamefont{Blake}},
  \bibinfo{author}{\bibfnamefont{A.~N.} \bibnamefont{Grigorenko}},
  \bibinfo{author}{\bibfnamefont{K.~S.} \bibnamefont{Novoselov}},
  \bibinfo{author}{\bibfnamefont{T.~J.} \bibnamefont{Booth}},
  \bibinfo{author}{\bibfnamefont{T.}~\bibnamefont{Stauber}},
  \bibinfo{author}{\bibfnamefont{N.~M.} \bibnamefont{Peres}}, \bibnamefont{and}
  \bibinfo{author}{\bibfnamefont{A.~K.} \bibnamefont{Geim}},
  \bibinfo{journal}{Science} \textbf{\bibinfo{volume}{320}},
  \bibinfo{pages}{1308} (\bibinfo{year}{2008}).

\bibitem[{\citenamefont{Dikin et~al.}(2007)\citenamefont{Dikin, Stankovich,
  Zimney, Piner, Dommett, Evmenenko, Nguyen, and Ruoff}}]{dikin2007preparation}
\bibinfo{author}{\bibfnamefont{D.~A.} \bibnamefont{Dikin}},
  \bibinfo{author}{\bibfnamefont{S.}~\bibnamefont{Stankovich}},
  \bibinfo{author}{\bibfnamefont{E.~J.} \bibnamefont{Zimney}},
  \bibinfo{author}{\bibfnamefont{R.~D.} \bibnamefont{Piner}},
  \bibinfo{author}{\bibfnamefont{G.~H.} \bibnamefont{Dommett}},
  \bibinfo{author}{\bibfnamefont{G.}~\bibnamefont{Evmenenko}},
  \bibinfo{author}{\bibfnamefont{S.~T.} \bibnamefont{Nguyen}},
  \bibnamefont{and} \bibinfo{author}{\bibfnamefont{R.~S.} \bibnamefont{Ruoff}},
  \bibinfo{journal}{Nature} \textbf{\bibinfo{volume}{448}},
  \bibinfo{pages}{457} (\bibinfo{year}{2007}).

\bibitem[{\citenamefont{Bonaccorso et~al.}(2010)\citenamefont{Bonaccorso, Sun,
  Hasan, and Ferrari}}]{bonaccorso2010graphene}
\bibinfo{author}{\bibfnamefont{F.}~\bibnamefont{Bonaccorso}},
  \bibinfo{author}{\bibfnamefont{Z.}~\bibnamefont{Sun}},
  \bibinfo{author}{\bibfnamefont{T.}~\bibnamefont{Hasan}}, \bibnamefont{and}
  \bibinfo{author}{\bibfnamefont{A.}~\bibnamefont{Ferrari}},
  \bibinfo{journal}{Nature photonics} \textbf{\bibinfo{volume}{4}},
  \bibinfo{pages}{611} (\bibinfo{year}{2010}).

\bibitem[{\citenamefont{Alisultanov and
  Reis}(2016{\natexlab{a}})}]{alisultanov2016quantum}
\bibinfo{author}{\bibfnamefont{Z.}~\bibnamefont{Alisultanov}} \bibnamefont{and}
  \bibinfo{author}{\bibfnamefont{M.}~\bibnamefont{Reis}}, \bibinfo{journal}{EPL
  (Europhysics Letters)} \textbf{\bibinfo{volume}{113}}, \bibinfo{pages}{28004}
  (\bibinfo{year}{2016}{\natexlab{a}}).

\bibitem[{\citenamefont{Alisultanov and
  Reis}(2016{\natexlab{b}})}]{alisultanov2016magneto}
\bibinfo{author}{\bibfnamefont{Z.}~\bibnamefont{Alisultanov}} \bibnamefont{and}
  \bibinfo{author}{\bibfnamefont{M.}~\bibnamefont{Reis}},
  \bibinfo{journal}{Physics Letters A} \textbf{\bibinfo{volume}{380}},
  \bibinfo{pages}{470} (\bibinfo{year}{2016}{\natexlab{b}}).

\bibitem[{\citenamefont{Reis and Soriano}(2013)}]{reis2013electrocaloric}
\bibinfo{author}{\bibfnamefont{M.}~\bibnamefont{Reis}} \bibnamefont{and}
  \bibinfo{author}{\bibfnamefont{S.}~\bibnamefont{Soriano}},
  \bibinfo{journal}{Applied Physics Letters} \textbf{\bibinfo{volume}{102}},
  \bibinfo{pages}{112903} (\bibinfo{year}{2013}).

\bibitem[{\citenamefont{Reis}(2012)}]{reis2012oscillating}
\bibinfo{author}{\bibfnamefont{M.}~\bibnamefont{Reis}},
  \bibinfo{journal}{Applied Physics Letters} \textbf{\bibinfo{volume}{101}},
  \bibinfo{pages}{222405} (\bibinfo{year}{2012}).

\bibitem[{\citenamefont{Otsuji et~al.}(2013)\citenamefont{Otsuji, Tombet,
  Satou, Ryzhii, and Ryzhii}}]{otsuji2013terahertz}
\bibinfo{author}{\bibfnamefont{T.}~\bibnamefont{Otsuji}},
  \bibinfo{author}{\bibfnamefont{S.~B.} \bibnamefont{Tombet}},
  \bibinfo{author}{\bibfnamefont{A.}~\bibnamefont{Satou}},
  \bibinfo{author}{\bibfnamefont{M.}~\bibnamefont{Ryzhii}}, \bibnamefont{and}
  \bibinfo{author}{\bibfnamefont{V.}~\bibnamefont{Ryzhii}},
  \bibinfo{journal}{IEEE Journal of Selected Topics in Quantum Electronics}
  \textbf{\bibinfo{volume}{19}}, \bibinfo{pages}{8400209}
  (\bibinfo{year}{2013}).

\bibitem[{\citenamefont{Xia et~al.}(2009)\citenamefont{Xia, Mueller, Lin,
  Valdes-Garcia, and Avouris}}]{xia2009ultrafast}
\bibinfo{author}{\bibfnamefont{F.}~\bibnamefont{Xia}},
  \bibinfo{author}{\bibfnamefont{T.}~\bibnamefont{Mueller}},
  \bibinfo{author}{\bibfnamefont{Y.-m.} \bibnamefont{Lin}},
  \bibinfo{author}{\bibfnamefont{A.}~\bibnamefont{Valdes-Garcia}},
  \bibnamefont{and} \bibinfo{author}{\bibfnamefont{P.}~\bibnamefont{Avouris}},
  \bibinfo{journal}{Nature nanotechnology} \textbf{\bibinfo{volume}{4}},
  \bibinfo{pages}{839} (\bibinfo{year}{2009}).

\bibitem[{\citenamefont{Matyba et~al.}(2010)\citenamefont{Matyba, Yamaguchi,
  Eda, Chhowalla, Edman, and Robinson}}]{matyba2010graphene}
\bibinfo{author}{\bibfnamefont{P.}~\bibnamefont{Matyba}},
  \bibinfo{author}{\bibfnamefont{H.}~\bibnamefont{Yamaguchi}},
  \bibinfo{author}{\bibfnamefont{G.}~\bibnamefont{Eda}},
  \bibinfo{author}{\bibfnamefont{M.}~\bibnamefont{Chhowalla}},
  \bibinfo{author}{\bibfnamefont{L.}~\bibnamefont{Edman}}, \bibnamefont{and}
  \bibinfo{author}{\bibfnamefont{N.~D.} \bibnamefont{Robinson}},
  \bibinfo{journal}{Acs Nano} \textbf{\bibinfo{volume}{4}},
  \bibinfo{pages}{637} (\bibinfo{year}{2010}).

\bibitem[{\citenamefont{Zhou et~al.}(2014)\citenamefont{Zhou, Xu, Fan, Qi, Li,
  Bai, and Ren}}]{zhou2014graphene}
\bibinfo{author}{\bibfnamefont{Y.}~\bibnamefont{Zhou}},
  \bibinfo{author}{\bibfnamefont{X.}~\bibnamefont{Xu}},
  \bibinfo{author}{\bibfnamefont{H.}~\bibnamefont{Fan}},
  \bibinfo{author}{\bibfnamefont{M.}~\bibnamefont{Qi}},
  \bibinfo{author}{\bibfnamefont{J.}~\bibnamefont{Li}},
  \bibinfo{author}{\bibfnamefont{J.}~\bibnamefont{Bai}}, \bibnamefont{and}
  \bibinfo{author}{\bibfnamefont{Z.}~\bibnamefont{Ren}},
  \bibinfo{journal}{Graphene Optoelectronics} pp. \bibinfo{pages}{209--234}
  (\bibinfo{year}{2014}).

\bibitem[{\citenamefont{Zhang et~al.}(2009)\citenamefont{Zhang, Tang, Girit,
  Hao, Martin, Zettl, Crommie, Shen, and Wang}}]{zhang2009direct}
\bibinfo{author}{\bibfnamefont{Y.}~\bibnamefont{Zhang}},
  \bibinfo{author}{\bibfnamefont{T.-T.} \bibnamefont{Tang}},
  \bibinfo{author}{\bibfnamefont{C.}~\bibnamefont{Girit}},
  \bibinfo{author}{\bibfnamefont{Z.}~\bibnamefont{Hao}},
  \bibinfo{author}{\bibfnamefont{M.~C.} \bibnamefont{Martin}},
  \bibinfo{author}{\bibfnamefont{A.}~\bibnamefont{Zettl}},
  \bibinfo{author}{\bibfnamefont{M.~F.} \bibnamefont{Crommie}},
  \bibinfo{author}{\bibfnamefont{Y.~R.} \bibnamefont{Shen}}, \bibnamefont{and}
  \bibinfo{author}{\bibfnamefont{F.}~\bibnamefont{Wang}},
  \bibinfo{journal}{Nature} \textbf{\bibinfo{volume}{459}},
  \bibinfo{pages}{820} (\bibinfo{year}{2009}).

\bibitem[{\citenamefont{Lam and Guo}(2014)}]{lam2014bandgap}
\bibinfo{author}{\bibfnamefont{K.-T.} \bibnamefont{Lam}} \bibnamefont{and}
  \bibinfo{author}{\bibfnamefont{J.}~\bibnamefont{Guo}},
  \bibinfo{journal}{Graphene Optoelectronics} pp. \bibinfo{pages}{149--166}
  (\bibinfo{year}{2014}).

\bibitem[{\citenamefont{Katsnelson and
  Kat?s?nel?son}(2012)}]{katsnelson2012graphene}
\bibinfo{author}{\bibfnamefont{M.~I.} \bibnamefont{Katsnelson}}
  \bibnamefont{and} \bibinfo{author}{\bibfnamefont{M.~I.}
  \bibnamefont{Kat?s?nel?son}}, \emph{\bibinfo{title}{Graphene: carbon in two
  dimensions}} (\bibinfo{publisher}{Cambridge University Press},
  \bibinfo{year}{2012}).

\bibitem[{\citenamefont{Alisultanov}(2016)}]{alisultanov2016electron}
\bibinfo{author}{\bibfnamefont{Z.~Z.} \bibnamefont{Alisultanov}},
  \bibinfo{journal}{JETP Letters} \textbf{\bibinfo{volume}{103}},
  \bibinfo{pages}{598} (\bibinfo{year}{2016}).

\bibitem[{\citenamefont{Mucha-Kruczy{\'n}ski
  et~al.}(2010)\citenamefont{Mucha-Kruczy{\'n}ski, McCann, and
  Fal'ko}}]{mucha2010electron}
\bibinfo{author}{\bibfnamefont{M.}~\bibnamefont{Mucha-Kruczy{\'n}ski}},
  \bibinfo{author}{\bibfnamefont{E.}~\bibnamefont{McCann}}, \bibnamefont{and}
  \bibinfo{author}{\bibfnamefont{V.~I.} \bibnamefont{Fal'ko}},
  \bibinfo{journal}{Semiconductor Science and Technology}
  \textbf{\bibinfo{volume}{25}}, \bibinfo{pages}{033001}
  (\bibinfo{year}{2010}).

\bibitem[{\citenamefont{McCann and Koshino}(2013)}]{mccann2013electronic}
\bibinfo{author}{\bibfnamefont{E.}~\bibnamefont{McCann}} \bibnamefont{and}
  \bibinfo{author}{\bibfnamefont{M.}~\bibnamefont{Koshino}},
  \bibinfo{journal}{Reports on Progress in Physics}
  \textbf{\bibinfo{volume}{76}}, \bibinfo{pages}{056503}
  (\bibinfo{year}{2013}).

\bibitem[{\citenamefont{Ohta et~al.}(2006)\citenamefont{Ohta, Bostwick,
  Seyller, Horn, and Rotenberg}}]{ohta2006controlling}
\bibinfo{author}{\bibfnamefont{T.}~\bibnamefont{Ohta}},
  \bibinfo{author}{\bibfnamefont{A.}~\bibnamefont{Bostwick}},
  \bibinfo{author}{\bibfnamefont{T.}~\bibnamefont{Seyller}},
  \bibinfo{author}{\bibfnamefont{K.}~\bibnamefont{Horn}}, \bibnamefont{and}
  \bibinfo{author}{\bibfnamefont{E.}~\bibnamefont{Rotenberg}},
  \bibinfo{journal}{Science} \textbf{\bibinfo{volume}{313}},
  \bibinfo{pages}{951} (\bibinfo{year}{2006}).

\bibitem[{\citenamefont{Guo et~al.}(2010)\citenamefont{Guo, Liu, Chen, Zhu,
  Fang, and Gong}}]{guo2010controllable}
\bibinfo{author}{\bibfnamefont{B.}~\bibnamefont{Guo}},
  \bibinfo{author}{\bibfnamefont{Q.}~\bibnamefont{Liu}},
  \bibinfo{author}{\bibfnamefont{E.}~\bibnamefont{Chen}},
  \bibinfo{author}{\bibfnamefont{H.}~\bibnamefont{Zhu}},
  \bibinfo{author}{\bibfnamefont{L.}~\bibnamefont{Fang}}, \bibnamefont{and}
  \bibinfo{author}{\bibfnamefont{J.~R.} \bibnamefont{Gong}},
  \bibinfo{journal}{Nano letters} \textbf{\bibinfo{volume}{10}},
  \bibinfo{pages}{4975} (\bibinfo{year}{2010}).

\bibitem[{\citenamefont{Dai and Yuan}(2010)}]{dai2010modulating}
\bibinfo{author}{\bibfnamefont{J.}~\bibnamefont{Dai}} \bibnamefont{and}
  \bibinfo{author}{\bibfnamefont{J.}~\bibnamefont{Yuan}},
  \bibinfo{journal}{Journal of Physics: Condensed Matter}
  \textbf{\bibinfo{volume}{22}}, \bibinfo{pages}{225501}
  (\bibinfo{year}{2010}).

\bibitem[{\citenamefont{Raza}(2012)}]{raza2012graphene}
\bibinfo{author}{\bibfnamefont{H.}~\bibnamefont{Raza}},
  \emph{\bibinfo{title}{Graphene nanoelectronics: metrology, synthesis,
  properties and applications}} (\bibinfo{publisher}{Springer Science \&
  Business Media}, \bibinfo{year}{2012}).

\end{thebibliography}
\end{document}